# Evaluation and thermodynamic optimization of phase diagram of lithium niobate tantalate solid solutions

Umar Bashir[a*], Detlef Klimm[a], Michael Rusing[b], Matthias Bickermann[a], Steffen Ganschow[a**]

[a]Leibniz-Institut für Kristallzüchtung (IKZ), Max-Born-Str. 2, 12489 Berlin, Germany

[b]Institute of Applied Physics, Technische Universität Dresden, Nöthnitzer Strasse 61, 01187 Dresden, Germany

[b]Paderborn University, Integrated Quantum Optics, Institute for Photonic Quantum Systems (PhoQS), Warburger Str. 100, 33098 Paderborn, Germany

Corresponding authours:

*umar.ganie@ikz-berlin.de

** steffen.ganschow@ikz-berlin.de

**Abstract**

The phase diagram of the lithium niobate – lithium tantalate solid solutions was investigated using experimental data from differential thermal analysis (DTA) and crystal growth. We used XRF analysis to determine the elemental composition of crystals. Based on the Neumann-Kopp rule, essential data of end members lithium niobate (LN) and lithium tantalate (LT) was created. The heats of fusion of end members given by DTA measurements of LN (103 kJ/mol at 1531 K) and LT (289 kJ/mol at 1913 K) were given as input parameters to generate the data. This data served as the basis for calculating a phase diagram for LN-LT solid solutions. Finally, based on the experimental data and thermodynamic solution model, the phase diagram was optimized in the Calphad Factsage module. We also generated thermodynamic parameters for Gibb's excess energy of the solid solution. A plot of segregation coefficient as a function of Ta concentration was derived from the phase diagram.

**Introduction**

In contemporary materials science, lithium tantalate (LiTaO3, LT) and lithium niobate (LiNbO3, LN) are two of the most researched oxide compounds. The various applications of these materials in functional electronics have generated curiosity about their utilization in thin films, micro- and nanopowders, single or polycrystalline crystals, and ceramics. Additionally, they may differ in their chemical composition, be congruent or stoichiometric, and include different dopants. Such a wide range of types and applications necessitates extra study to adjust and improve material characteristics [1, 2].

$LiNb_{1-x}Ta_xO_3$ (LNT) solid solutions have lately received attention, and they present opportunities for combining the benefits of the two materials. According to the literature, they form solid solutions over the whole composition range. And by this way, the physical properties of materials, particularly ferroelectric and nonlinear optical properties, can be tuned as a function of composition. For instance, tuning the birefringence and refractive index as a Ta/Nb ratio function. A previous study shows zero birefringence in a composition range of $0.93 \leq x \leq 0.94$ at 293K [3, 4]. Also, an optically isotropic crystal (a crystal with zero birefringence) maintains the unique benefits of ferroelectrics, including substantial electro-optic and nonlinear optical coefficients [5]. The birefringence of LNT mixed crystals, was tested by Wood et al. for various compositions and temperatures, and they discovered that it constantly changed up to the ferroelectric Curie temperature Tc. In particular, it is possible to anticipate that the LNT solid solutions will display strong piezoelectric coefficients (comparable to LN) and at the same time LT features that are stable at high temperatures [4, 6, 7].

The solid solutions of LNT may serve as the foundation for ceramic materials with ferroelectric, superionic, semiconductor, and even mixed properties. As a result, this quasi-binary material system is extensively investigated using various techniques [8]. The LNT solid solutions are also appealing as a model system of how composition affects the structural properties of mixed oxides.

LN and LT endmembers are isostructural, have comparable lattice and atomic positional parameters, and have similar ionic radii and valence states for Ta and Nb. However, the separation between solidus and liquidus lines in the LN-LT phase diagram is large [9, 10]. This makes the growth of homogeneous single crystals of LNT solid solution difficult. The melt around the crystal gets increasingly LN rich as LN is repeatedly rejected from the solid-liquid interface during crystal growth, this causes compositional inhomogeneity in the crystal along the growth direction e.g. using the Czochralski (CZ) method [4, 5, 11-13]. In contrast, crystals with uniform Ta and Nb composition have been grown using an edge-defined film-fed growth method [11]. Still, these crystals were not high-quality showing bubbles, microcracks, and sub grain boundaries.

Several studies have reported that the Ta/Nb ratio in LNT affects structural characteristics such as lattice parameters, phase transition temperature etc. [4, 9-12]. Wood et al. reported absolute values of LNT solutions' birefringence [4], and Shimura et al. [5] published the refractive indices of Ta-rich LNT crystals. Yun-Cheng Ge et al. studied ferroelectric phase transition using Raman analysis [14]. Bartasyte et al. [3] studied the effect of Ta/Nb compositional variation on lattice parameters. Roshchupkin et al. [7] studied the crystal for sensor and actuator applications. In our other work, we observed the variation of Curie temperature ($T_c$) with Ta/Nb covering the whole composition range [15].

Despite its importance, the LN-LT phase diagram has not been studied in detail. The first study on the phase diagram was conducted by Peterson et al. [9, 10]. Crystals were produced by heating the powder mixtures of LNT in a platinum/iridium crucible and allowed to crystallize by cooling. This study was carried out on LN/LT congruent crystals, where the liquidus and solidus meet at the end of LN and LT in the phase diagram. However, in their following study [10], they used stoichiometric LN and LT and observed a different behavior of the phase diagram where the *liquidus and* solidus *do not meet at the end members of LN and LT*. This is obvious because the stoichiometric and congruent melting compositions do not coincide; hence, the connection between the stoichiometric compositions rather represents an isopleth section through the $Li_2O$-$Nb_2O_5$-$Ta_2O_5$ ternary system. Therefore, it is important to re-investigate the phase diagram on LN-LT system over the whole composition ranges ($x = 0 \, to \, x = 1$) between the congruent compositions, to obtain a truly quasi-binary diagram. We are particularly interested in the mixed crystal's growth, miscibility, and solidus-liquidus separation. In this work, we will also construct the phase diagram based on a thermodynamic model where we consider the excess enthalpies of mixing.

Furthermore, LNT single crystal specimens used in this study were taken from large volume crystals which are in the size of 5 cm in length and 1.5 cm in diameter. The previous LNT single crystals grown by Bartasyte et al. [3] had a length of 1.5 cm, and Roshchupkin et al. [7] grew crystals measuring 2 cm in length and diameter.

**Experiment**

Single crystals were grown from LNT melts using the Czochralski process with induction heating. A cylindrical crucible made from platinum (for niobium-rich compositions with low melting temperature) or iridium (for moderate to high tantalum contents) measuring 60 mm in height and 60 mm in diameter was employed. The latter process was conducted in a protective atmosphere, namely argon with a small (< 1 vol%) addition of oxygen, otherwise in air. Starting materials were mixtures of the congruently melting LN, respectively LT, prepared from lithium carbonate ($Li_2CO_3$, Alfa Aesar, 5N), niobium pentoxide ($Nb_2O_5$, H.C. Stark, 4N5), and tantalum pentoxide ($Ta_2O_5$, Fox Chemicals, 4N), all dried before use. The congruent compositions of LN and LT are at 48.38% $Li_2O$ [11] and 48.46% $Li_2O$ [12], respectively. We took the average 48.4 mol% $Li_2O$ for all compositions. All crystals were grown along the c-axis using LN, LT or LNT seeds, respectively, at 0.5 mm/h pulling rate for all compositions.

Energy dispersive micro X-ray fluorescence (µ-XRF) measurements determined the elemental (Nb, Ta) distribution in the grown single crystals. The measurements were carried out with a Bruker M4 TORNADO spectrometer at a low-pressure environment (20 mbar). The X-ray source was a tube with Rh anode working at 50 kV and 200 µA. Bremsstrahlung was focused at the sample surface using polycapillary X-ray optics, yielding a high spatial resolution (beam width) of about 20 µm, and the measurement time per pixel was 20 ms. The fluorescence signal was detected with a circular silicon drift detector. Quantification was done by the built-in routines of the spectrometer using the fundamental parameters database.

A NETZSCH STA 449C "Jupiter" thermal analyzer was used to perform simultaneous thermogravimetry (TG) and differential thermal analysis (DTA). A DTA/TG sample holder outfitted with platinum crucibles, Pt/Pt90Rh10 thermocouples, and a flowing mixture of 20 ml/min Ar and 20 ml/min $O_2$ enabled readings up to 1923 K. Figure 1a and 1b shows the size of crystals used for DTA measurements.

For the thermodynamic assessment of the phase diagram, FactSage 8/8.2 was used. The JANAF [16-18] table does not contain the data for LN and LT end members. Thus, the thermodynamic data of LN and LT was calculated from the Fact-PS database from the $Li_2O+Nb_2O_5$ or ($Ta_2O_5$) data, and both compounds were modified using measured Cp data [19, 20].

**Results and discussion**

Figure 1 shows the DTA results for LN and LT end members and for LNT solid solutions of different compositions. The onset of DTA peaks corresponds to the melting point and this temperature will be taken as a *solidus* in the optimization of the phase diagram. The temperature, at which the DTA curve returns to the baseline after the melting represents the

*liquidus* in the phase diagram. The DTA measurements were taken twice, first heating corresponds to the heating of as-grown crystal. After the first heating, the crystal melts and recrystallizes upon cooling. The second heating is carried out on the recrystallized sample. As can be seen from figure 1a the onset is sharp for end members ($x = 1$) in both first and second heating. For pure LT, the liquidus temperature is not visible in the DTA curve. This is because of the temperature limit of the experimental equipment. Therefore, for pure LT the melting temperature is taken from literature [15]. Moreover, the end members do not show any difference in melting and solidification temperatures for first and second heating. This is obvious because there is no component segregation in the end members.

Figure 1a shows that the melting temperature of the solid solution increases with increasing LT fraction. This is the expected behavior since the melting point of LT is 400 K higher than that of LN. However, width of the melting peaks and the separation between solidus and liquidus temperatures varies with composition. For solid solutions, melting extends over a significantly larger temperature range than in case of the end members, with a maximum for compositions around .??? The separation between solidus and liquidus decreases as the solid solution approaches one of the endmembers, i.e. for Nb or Ta rich solid solutions. In the second heating, melting is even more prolonged in solid solutions. This is attributed to segregation that occurs during solidification, leading to more inhomogeneous specimen containing LN and LT rich regions. As a result, in the second heating Nb rich regions will start melting at lower temperatures and slowly solve the remaining solid. This tendency is observed for the second heating of all compositions of solid solutions. For??? , the second heating shows the melting at decreased temperature (1876 K) as compared to first heating (1890 K) and liquidus temperature for this composition cannot be clearly identified as shown in the DTA curve. The heat absorbed in the endothermic process of melting is a function of composition and mass of the sample. Therefore, the peak area is different for different compositions.

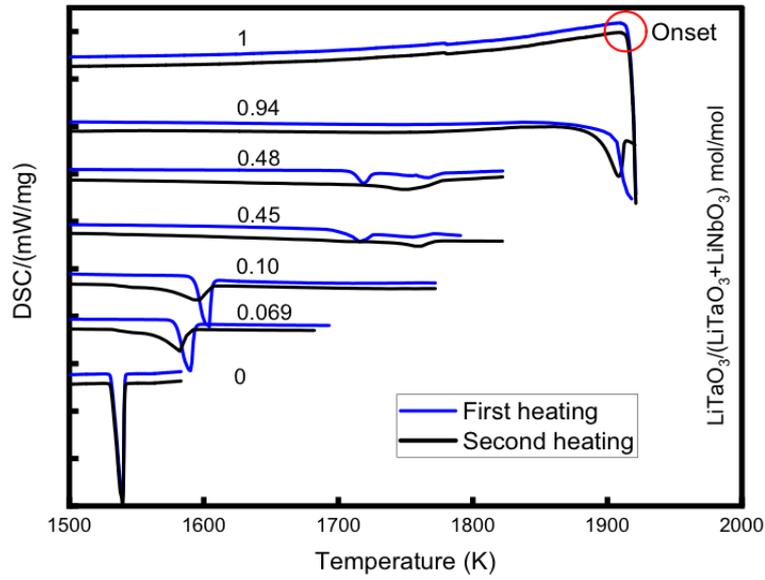

(a)

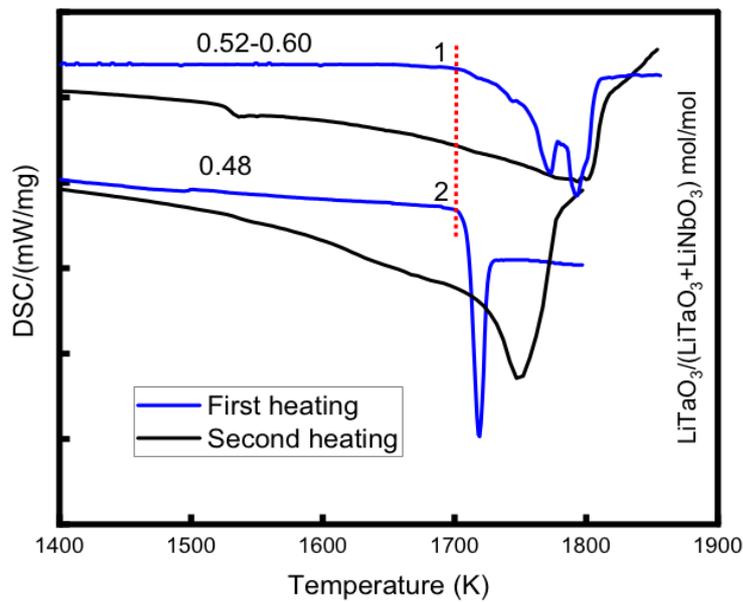

(b)

Figure 1. (a) DTA plots of LN, LT and LNT solid solutions with first heatings (blue curve) and second heatings (black curve). The legend on the right side of the y-axis shows the molar ratio.

(b) DTA plots of sample represented by Figure 1b (specimen 1, $x$ = 0.52-0.60) shows no clear onset but prolonged melting in both heating cycles. DTA plots corresponding to Figure 1b

(specimen 2, $x$ = 0.48) show clear melting temperature in first heating but a prolonged melting in the second heating.

Figure 1b shows DTA curves for two selected samples with comparable compositions. It is important to note that the first heating of sample 1 shows highly inhomogeneous regions and a prolonged melting. The Ta and Nb elemental agglomeration are clearly visible e.g. in µXRF maps, as shown exemplarily in Figure 2b. Although the composition of the sample in Figure 2b varies from ($x$ = 0.20-0.60), the major part of the crystal is dominated by Ta rich composition, however the small Nb rich segregated regions will cause a prolonged melting. On the other hand, the first heating of sample 2 in Figure 1b shows a sharp melting at nearly the same temperature as sample 1 (red dotted line in Figure 1b). The reason for this sharp melting is that the sample is less inhomogeneous, comparable to a sample that is depicted in figure 2a. Here the compositional variation is ($x$ = 0.46-0.51) with no pronounced agglomeration such as in figure 2b. In the second heating the onset temperature could not be identified. Since the composition of two samples is similar, they show a similar inhomogeneity after the first melting and resolidification as can be seen from prolonged melting behavior in both samples (black curves), but during subsequent heating, the partial pre-melting caused by segregation will lead to different DTA curves. Therefore, for the optimization of phase diagram, it is necessary to take the solidus and liquidus temperatures determined from the first heating. Table 1 lists the solidus and liquidus temperatures of homogeneous crystals as represented by Figure 1a. The solid solutions show some variation of solidus and liquidus temperatures which are indicated by error ranges. The error was estimated with respect to the variation in composition in the samples. It is important to note here that the specimens used in Figure 1b are not the same as shown in Figure 2. The µXRF maps as shown in Figure 2 are representative images of comparable compositions used in Figure 1b. Here, the important point is to define inhomogeneities in the crystal. As can be seen the gradient (Δx) of x in Figure 2a is Δx ≤ 5 ($x$ = 0.46-0.51). On the other hand, for Figure 2b this gradient is quite large Δx = 40 ($x$ = 0.20-0.60). For all other crystals grown at IKZ the composition gradient is shown in Table 1. It can be seen that for some samples, Δx is less than 1. Therefore, crystals grown at IKZ show very high level of homogeneity as compared to some previous studies as referred earlier.

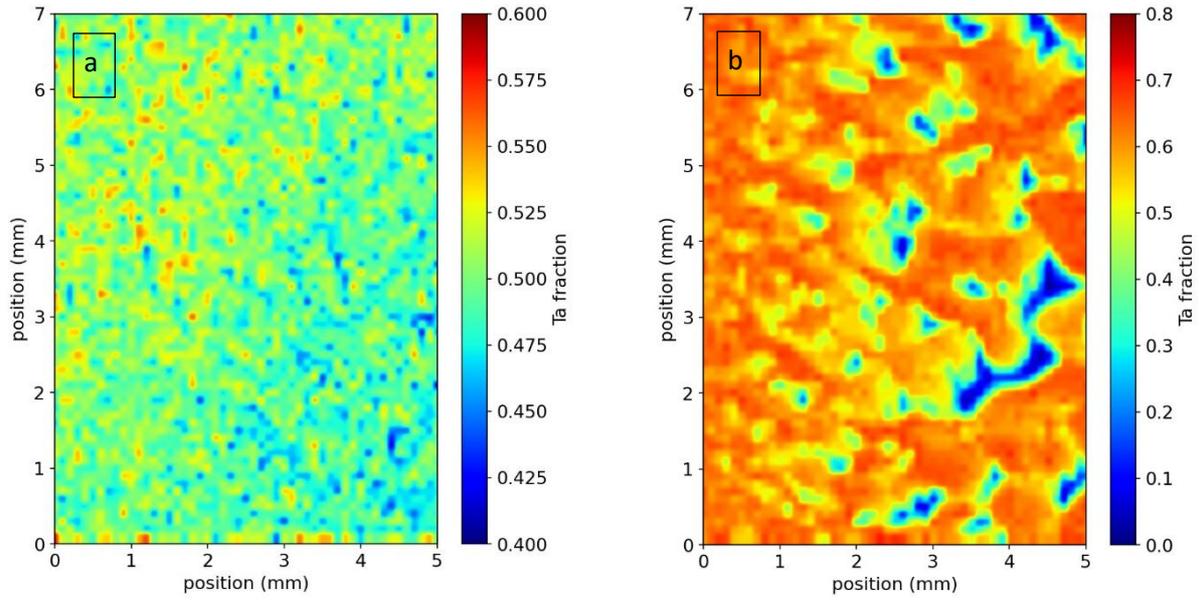

Figure 2. XRF elemental mapping of LNT mixed crystals (z-cut). a). Less segregated crystal (homogeneous). b). Crystal with higher segregation between Ta and Nb (inhomogeneous).

Table 1. Crystal composition and corresponding solidus and liquidus temperatures from Figure 1a

| LT mole fraction | Solidus Temperature (K) | Liquidus Temperature (K) |
| --- | --- | --- |
| 0 | 1523 | 1523 |
| 0.069 | 1550 ±10 | 1582 ±10 |
| 0.10 ($\Delta x < 1\%$) | 1570 ±5 | 1615 ±5 |
| 0.445 - 0.468 ($\Delta x = 2.3\%$) | 1670 ±10 | 1780 ±10 |
| 0.48 | 1690 ±10 | 1780 ±10 |
| 0.941-0.942 ($\Delta x < 1\%$) | 1887 ±5 | 1909 ±5 |
| 1 | 1917 | 1917 |

**Thermodynamic model**

Before proceeding with a solution model for the phase diagram calculation, the thermodynamic parameters for LN and LT end members are required in the database. This database was used to determine the thermodynamic parameters (specific heat capacity ($Cp$), standard enthalpy ($\Delta H$), and entropy ($\Delta S$)) of LiNbO$_3$ and LiTaO$_3$ based on Neumann-Kopp (NK) rule from the Factsage mixer module [5]. As per Neumann-Kopp rule, the molar heat capacity ($Cp(T)$) of a compound is calculated by adding, at a specific temperature, the molar heat capacities of its individual components, each multiplied by their respective quantities within the compound [21]. A private database was thus created in the Factsage "Compound module", containing all relevant compounds and mixtures:

$$2LiNbO_3 = Nb_2O_5 + Li_2O$$

$$2LiTaO_3 = Ta_2O_5 + Li_2O$$

Table 2 lists the thermodynamic parameters of LN and LT obtained from the above process. To make the phases stable in the Li$_2$O-(Nb,Ta)$_2$O$_5$ system, the initial values of $H^o_{298}$ ($\Delta H_{LN}$ = -1080 kJ/mole, $\Delta H_{LT}$ = -1140 kJ/mole) were corrected to final values listed in Table 2. The correction is necessary because the sole application of the NK rule disregards the formation enthalpy of the complex oxides LN and LT from the simple oxides Li$_2$O, Nb$_2$O$_5$ and Ta$_2$O$_5$, respectively.

Since the end members are lithium niobate and lithium tantalate, the following reaction is considered based on Hess law [18]:

$$LiNb_{1-x}Ta_xO_3 = xLiTaO_3 + (1-x)LiNbO_3$$

The Gibbs excess energy G$^{excess}$ for LN-LT mixtures is assumed to be relevant only in the solid phase. This is, because Nb and Ta possess very similar chemical properties, and the ionic radii of Nb$^{5+}$ and Ta$^{5+}$ in octahedral coordination are almost identical (78 pm). Indeed, it turned out, that already the assessment of the solid mixture phase could reproduce the experimental results satisfactory, if the liquid phase was treated ideal.

Many simple models exist for solid solutions of iso-structural compounds. Although these models are for mineralogical systems the assumptions on which the models are based are similar to the ones developed in this study. For example, Wood et al. uses the so-called Bragg-Williams model, which considers the substitution of Si by Al in Mg$_3$Al$_2$Si$_3$O$_{12}$ system [22, 23]. In this model complete mixing of end members is considered, and the charge balance is maintained during the cationic substitution. Therefore, this model is also applicable to our system. Another model is used by Ringwood et al. which considers the substitution of $Fe^{+2}$ by $Mg^{+2}$ in a $Fe_2SiO_4$—$Mg_2SiO_4$ system [24]. This model also considers complete cationic exchange while maintaining the charge balance. For several other binary solid solutions, Holland et.al [20] discuss seven models. The model used in this study is also based on complete mixing of end members. The assumptions in our model are the following: all octahedral distortion is disregarded, lithium oxide is kept constant throughout, complete correlation with $Nb^{+5}$-$Ta^{+5}$ occupancy site, this preserves the charge balance [19].

**Optimization of phase diagram:**

The experimental results of specific heat capacity are discussed in detail in our previous work [15]. Plot of specific heat capacities of LN and LT is shown in Figure 3. The plots compare the specific heat capacity calculated from Neumann-Kopp rule with experimental results and show they are in reasonable agreement. The agreement between calculated and experimental data is a first proof that the end members created from the mixer module are reliable and can be used to proceed for the creation of a solid solution compound module. However, it is important to note that only the experimental results will be used during the optimization of phase diagram [25]. A polynomial expression (Equation 1) was used to fit the experimental data:

$$C_p = a + b.T + C.T^{-2} + d.T^2 + e.T^{-1} + f.\sqrt{T} \qquad (1)$$

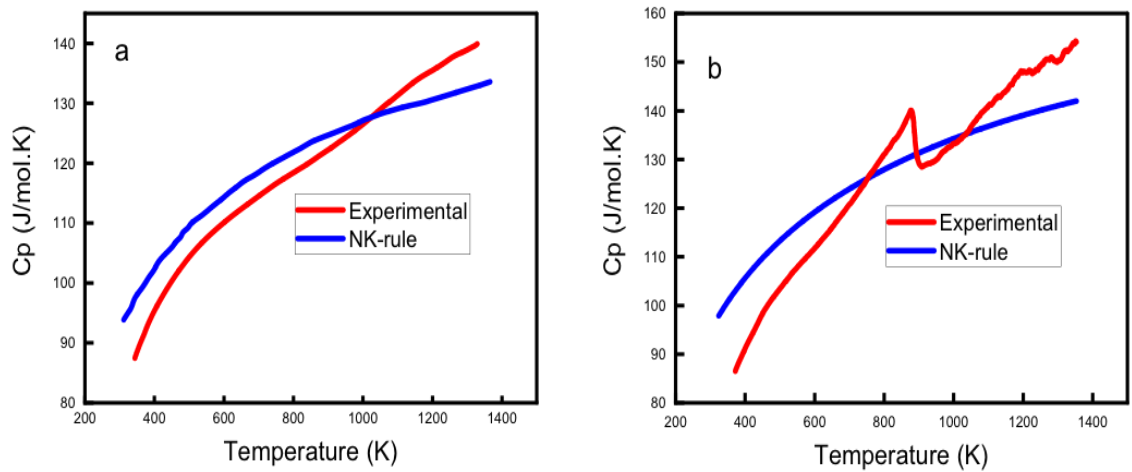

Figure 3. Specific heat capacity of a) $LiNbO_3$ and b) $LiTaO_3$ calculated from the thermodynamic database of elemental oxides (Neumann-Kopp rule) and experimental results

of specific heat capacity. The ferroelectric transition of LiNbO3 lies outside the temperature measurement range.

The CALPHAD method illustrates the Gibbs free energy composition dependent on a polynomial expansion. Therefore, in Factsage, the Redlich and Kister polynomial expansion is used [26] for our binary system.

$$G^{excess} = (1-x)x \sum_{J=0}^{n} Lj((1-x)x)^j \qquad (2)$$

Under constant temperature and pressure, The Redlich-Kister polynomial is used to model the excess molar Gibbs free energy in the context of solution thermodynamics, particularly in the study of non-ideal solutions. The polynomial is often employed to describe the composition dependence of Gibbs free energy in binary systems, *Lj* are the interaction coefficients determined through regression analysis based on experimental data, and *j* is the order of interaction. The Redlich-Kister polynomial provides a flexible means of representing the non-ideality in solution. The specific form of the Redlich-Kister polynomial may vary depending on the order chosen and the particular details of the system under consideration. Higher-order terms in the polynomial can capture more complex deviations from ideal behavior [27]. The excess enthalpy and entropy contributions, specific heat capacity of end members, heats of fusion of end members, experimental results from crystal growth and DTA are all input parameters given to the model to proceed for calculations in Factsage. The heats of fusion of end members is given by DTA measurements of LN (103 kJ/mol at 1531 K) and LT (289 kJ/mol at 1913 K).

Figure 4. shows the phase diagram optimized in the FactSage's "Phase Diagram" module. The shape of this pseudo-binary phase diagram is similar to the phase diagram determined by Peterson et al. [9]. Peterson's phase diagram is based on crystal growth experiments, it shows a large separation between solidus and liquidus indicating large segregation and his work does not provide any thermodynamic modeling of the phase diagram.

The solidus and liquidus temperatures were taken from the first heating of DTA experimental results. Data points with errors are shown with error bars in the phase diagram accounting for the fact that the solidus and liquidus temperatures are not sharp onsets in all cases. On the other hand, the errors in composition are hardly beyond 2.3 % (Table 1), as the DTA samples are prepared from homogeneous areas of crystals with a well-known composition. In contrast, melt and solid compositions from crystal growth are considered to be located at the liquidus in the phase diagram and at its corresponding solid composition along an isotherm. The graph shows that the crystal growth solid compositions lie within the equilibrium region. This coincides with the well-known fact that the effective segregation during crystal growth is always closer to unity than the segregation in equilibrium. Note that crystal growth experiments comprised three compositions without DTA results. Here, the solid and liquid composition from XRF results and melt, respectively, they were fitted to the solidus and liquidus lines on the phase diagram. The optimized phase diagram was then constructed by the FactSage "CALPHAD" module.

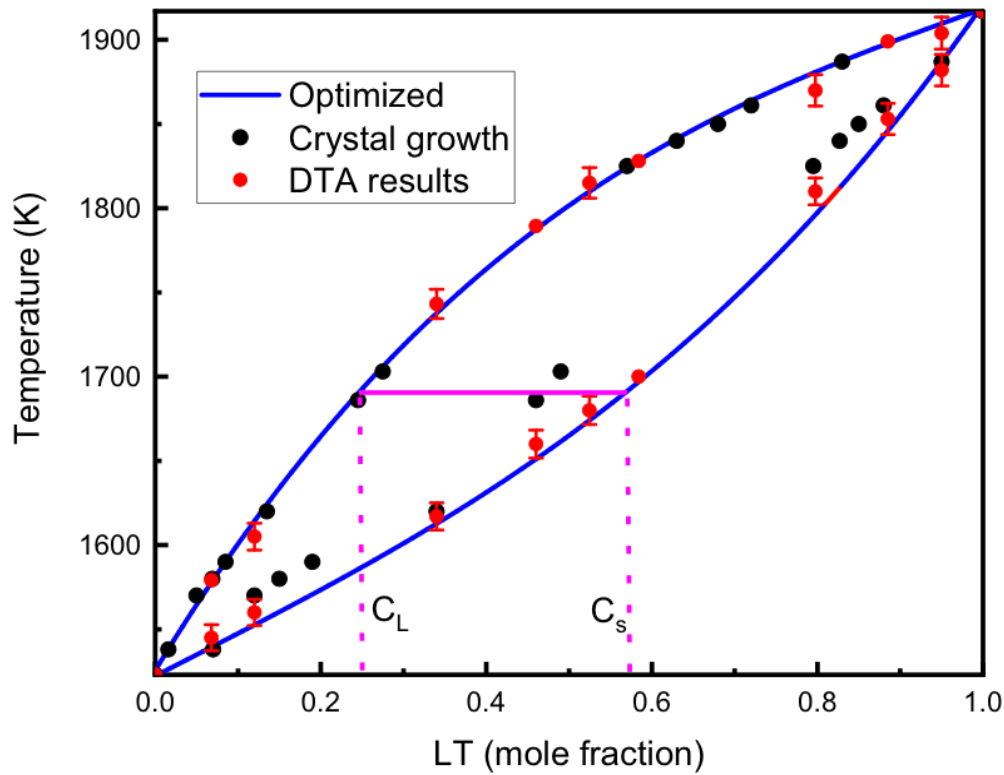

Figure 4. Optimized phase diagram based on experimental results of DTA (composition and corresponding solidus and liquidus temperature) as well as specific heat capacity and heats of fusion of end members (Blue curve). Crystal growth experimental data (black dots) and DTA results (red dots) also included the phase diagram.

Table 2. Thermodynamic parameters of end members calculated from the database using Neumann-Kop rule.

| Compound | $H^0_{298.15}$ (kJ/mol) | $S^0_{298.15}$ (J/mol.K) | $Cp$ (J/mol.K) |
|---|---|---|---|
| $LiNbO_3(s)$ | $-1256.000.00$ | 87.016 | $56.016 + 0.101.T - 2136090.4.T^{-2} - 3.903E-5.T^2 + 5172.367.T^{-1} + 295.42.T^{-0.5}$ (298–800K) $104.062 + 3.531.T - 5108494.4.T^{-2} - 8.537E-6.T^2 + 5172.3676.T^{-1}$ (800–1500K) |
| $LiTaO_3(s)$ | $-1336.78800$ | 90.502 | $166.127 + 0.0164.T - 388262.64.T^{-2} - 4.457E-6.T^2 + 5172.367.T^{-1} - 1536.79.T^{-0.5}$ (298–1843K) |

From the phase diagram, the values of standard enthalpies corresponding to $L_j$ coefficients of equation 2 are extracted. Using the coefficients and Ta/Nb compositions in equation 2, a plot of $G^{excess}$ (red curve) of the LNT solid solution is obtained as shown in figure 5. For end members, $G^{excess}$ is zero as there is no interaction involved. For a solid solution, the excess energy of mixing is negative (exothermic reaction) with the minimum at $x = 0.5$ (-2703 J/mole). Figure 5 also shows the plot for $G^{ideal}$ (black curve) with a minimum at $x = 0.50$ (-1725.9 J/mole). From the two curves, it is clear that $G^{excess}$ is of higher magnitude than $G^{ideal}$. Figure 5 corresponds to the sum of $L_0$ (-10$^4$) J/mole, $L_1$ (-30$^3$) J/mole and $L_2$ (-10$^3$) J/mole. While the shape of the curves is symmetric for all the three coefficients $L_0$, $L_1$ and $L_2$, but $L_0$ shows the dominant effect.

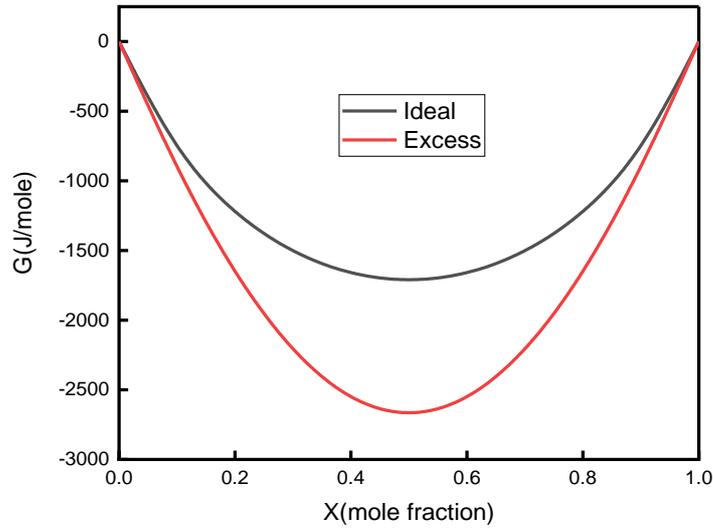

Figure 5. A room temperature plot for $G^{excess}$ and $G^{ideal}$ for LNT solid solution.

The phase diagram is used to calculate the segregation coefficient ($K$). $K$ is defined as the ratio of Ta composition in solid ($C_s$) to Ta composition in the melt ($C_l$).

$$K = C_s/C_l \tag{3}$$

From the phase diagram ($C_s$) composition corresponds to any point on the solidus. The corresponding melt composition can be determined by drawing an isotherm (horizontal tie line) on the liquidus ($C_l$) as shown in Figure 4 (solid magenta line connecting solidus and liquidus). The vertical projection from the end points of the horizontal line on x-axis corresponds to crystal and melt compositions (vertical dashed lines in Figure 4). The results of $K$ are plotted in figure 6. As defined by eq. 3, the segregation coefficient approaches unity as the LNT composition approaches LT. For any composition, the Ta segregation coefficient is always higher than unity. Therefore, the melt must get Ta deficient as the crystal grows.

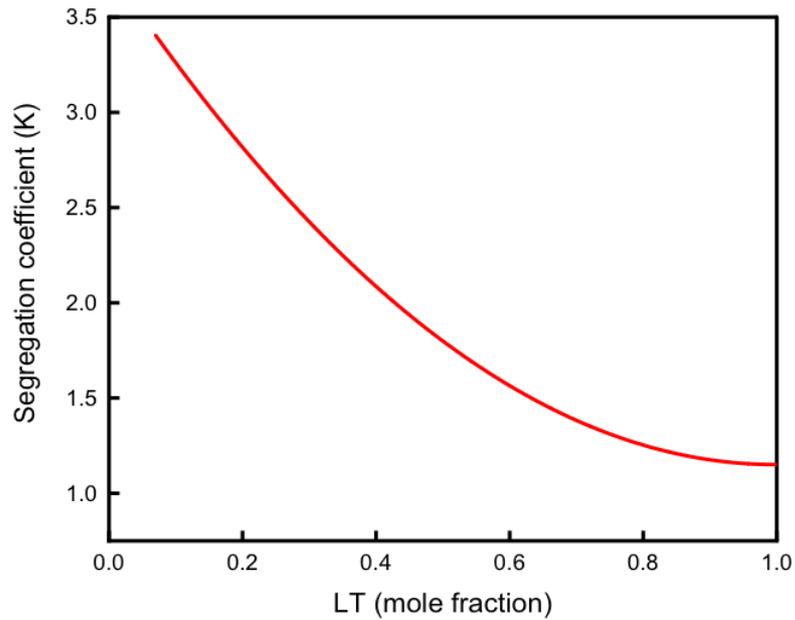

Figure 6. LT Segregation coefficient as a function of Ta mole fraction in the melt.

From the segregation coefficient, melt and solid composition during the solidification process can be determined using the Scheil Equation [28], when complete mixing the melt is assumed [29]:

$$Cs = KVo(1-Vs)^{K-1} \qquad (4)$$

Where *Cs* is the concentration of solute in the solid, *Vs* is the volume of a solidified fraction of the melt, and *Vo* is the volume of the liquid when growth commences. The Scheil plot in Figure 7 for solidification of a LNT melt with x = 0.5 shows that as a consequence from the segregation the initial Ta composition in the crystal is as high as = 0.85 (red arrow), but decreases in the melt (slope of blue curve) as well as in the solid (green curve) as solidification progresses and the liquidus temperature ($x_{melt}$ = 0.50, T = 1822 K) decreases to finally approach the growth temperature of pure LN (1523 K). Close to the latter point, LT is completely depleted in the melt, and thus the very bottom of the crystal (when all melt is solidified) will also contain no Ta. The inevitable segregation effects must be considered and accounted for to grow crystals that are useful application purposes.

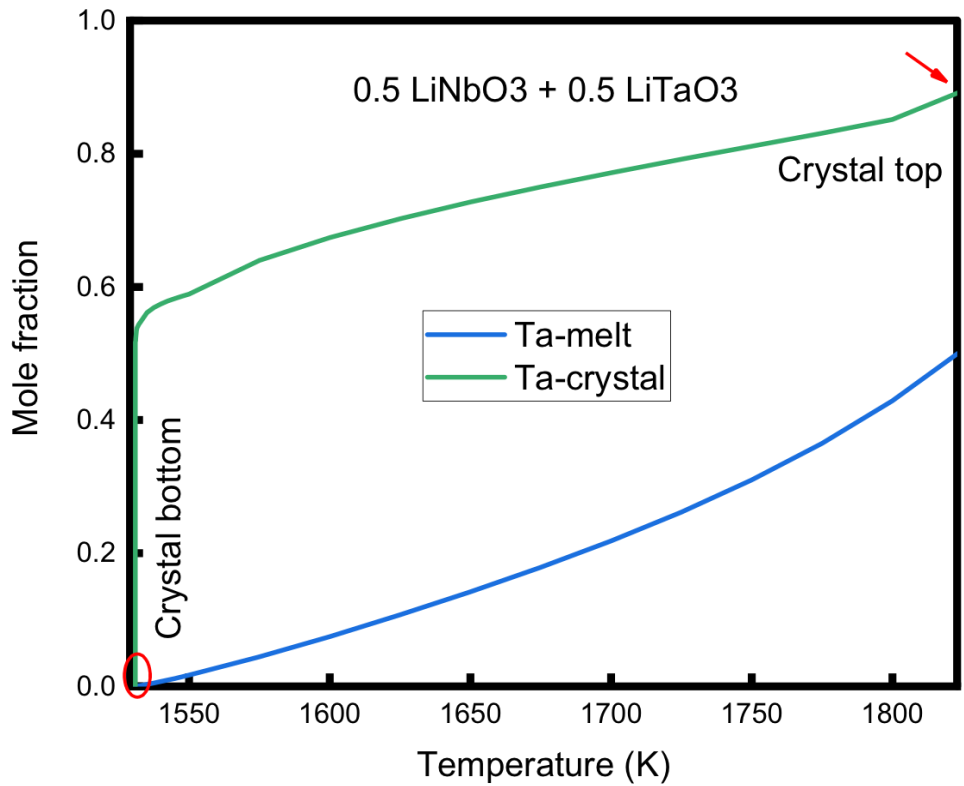

Figure 7. Scheil cooling of LNT solid solution with starting composition of x = 0.5. Blue curve: Ta composition decreases fast in the melt. Green curve: In the crystal, initial Ta content is higher, but also drops along growth axis given by the slope of the curve

**Conclusion:**

The LNT solid solutions were investigated with DTA and a detailed phase diagram is proposed based on a thermodynamic solution model. The phase diagram shows complete miscibility over all compositions $(x = 0 - x = 1)$ and a narrower separation between solidus and liquidus as compared to the previously studied phase diagram. A plot of the Gibbs excess energy shows significant deviation from ideality. The segregation coefficient of Ta calculated from the phase diagram is greater than unity for all compositions. The differences in composition upon solidification (e.g., Scheil cooling) must be considered to obtain useful and optimized crystals for the intended application.

**Acknowledgments**

This work was supported by Deutsche Forschungsgemeinschaft (DFG) in the framework of Research Group FOR5044 "Periodic low-dimensional defect structures in polar oxides", grant no. 426703838.